\documentclass[%
 reprint,
groupedaddress,
 amsmath,amssymb,
 aps,
prb,
]{revtex4-1}

\usepackage{graphicx}
\usepackage{dcolumn}
\usepackage{bm}

\begin{document}

\newcommand{\be}{\begin{equation}}
\newcommand{\ee}[1]{\label{#1}\end{equation}}
\newcommand{\bem}{\begin{eqnarray}}
\newcommand{\eem}[1]{\label{#1}\end{eqnarray}}
\newcommand{\eq}[1]{Eq.~(\ref{#1})}
\newcommand{\Eq}[1]{Equation~(\ref{#1})}
\newcommand{\vp}[2]{[\mathbf{#1} \times \mathbf{#2}]}


\title{ Spin and mass  superfluidity in ferromagnetic spin-1 Bose-Einstein condensate}

\author{E.  B. Sonin}
 \affiliation{Racah Institute of Physics, Hebrew University of
Jerusalem, Givat Ram, Jerusalem 91904, Israel}

\date{\today}

\begin{abstract}
The paper investigates the coexistence and interplay of spin and mass superfluidity in a ferromagnetic spin-1 Bose-Einstein condensate. Superfluidity is possible only in the presence of uniaxial anisotropy (linear and quadratic Zeeman effect). This follows  from the topology of the order-parameter space (vacuum manifold). According to the Landau criterion, the critical phase gradients, both for mass and spin supercurrents, vanish at the phase transition from the easy-plane to the easy-axis anisotropy. However, mass superfluidity is still possible at the phase transition. This is because the Landau criterion signals instability only with respect to nonsingular vortices with special ratio between circulations of mass and spin currents. Phase slips produced by these vortices are not sufficient for complete decay of supercurrents. Full decay of supercurrents requires phase slips with vortices of another topological class and larger energy.  These phase slips are suppressed by energetic barriers up to the upper critical velocity  (gradient) exceeding the Landau critical velocity. The upper critical velocity does not vanish nor has any anomaly in the critical point at the phase transition from the easy-plane to the easy-axis anisotropy.
\end{abstract}

\maketitle


\section{Introduction} \label{Intr} 

Spin superfluidity  was suggested in the 1970s \cite{ES-78b,ES-82} on the basis of analogy of magnetodynamics with hydrodynamics of superfluids.  Manifestation of spin superfluidity is a stable spin supercurrent proportional to the  gradient of  the spin phase (spin rotation angle in a plane) and  not accompanied by dissipation, in contrast to a dissipative spin diffusion current proportional to the gradient of spin density. Spin superfluidity was investigated for various magnetically ordered systems including superfluid $^3$He and solid ferro- and antiferromagnets (see recent reviews.\cite{Adv,Mac}) Recently they started to explore this phenomenon in spin-1 Bose-Einstein condensate (BEC).\cite{Duine,kim}

We define the term {\em superfluidity (superconductivity)} in its original meaning known from the times of Kamerlingh Ohnes and Kapitza: transport of some physical quantity (mass, charge, or spin) on macroscopical distances without essential dissipation. This does not mean that sources of dissipation are very weak or absent. A metal at zero temperature and without any defect has zero resistance, but it is not a superconductor. If a current flows without dissipation in a metal with defects at finite temperature, it has superconductivity.  Confusion on this issue is discussed in the review.\cite{Adv}

Superfluidity emerges if current states are metastable, i.e., any perturbation of the current state increases its energy. Then any elementary process of dissipation must overcome an energetic barrier, via either thermal activation, or quantum tunneling.  Metastability is conditioned by special topology of the order-parameter space.  Any point of this space is one from many degenerate ground states. If a closed streamline of a current in the configurational space maps on a path in   the order-parameter space, which cannot be reduced by continuous deformation (homotopy) to a single point one may expect current metastability.  For spin superfluidity this requires the easy-plane anisotropy for the spontaneous magnetization in ferromagnets or the antiferromagnetic vector in antiferromagnets.\cite{Adv,Mac} This anisotropy can emerge not only from crystal anisotropy, but also from long-range magnetostatic (dipolar) interaction as shown for the magnon condensate in  yttrium-iron-garnet  magnetic films.\cite{Son17}

The topological analysis yields a good qualitative hint when one may expect persistent currents, but  quantitative judgment on the possibility to observe these currents must be based on a direct check of metastability of current states. In the topological analysis it was supposed that gradients of phases (velocities) were very small, while at growing gradients one reaches the critical gradient values when the gradient kinetic energy becomes equal to the energy providing stability of current states. The first quantitative estimation of the supercurrent metastability was done by Landau for mass superfluidity (the famous Landau criterion). It was extended  on spin superfluidity.\cite{ES-78b,ES-82,Adv}  The Landau criterion for metastability is that any possible weak perturbation of the current state increases its energy. 

Although the Landau criterion signals when the current state ceases to be metastable, it does not answer the question of how can the current decay. Relaxation of the supercurrent is possible only via {\em phase slips} as was shown by  \citet{And6} long ago. A phase slip is a process, in which a vortex with a $2\pi$ phase variation around it crosses streamlines of the supercurrent. Any phase slip decreases the total phase variation across streamlines by $2\pi$. Phase slips are suppressed by energetic barriers. The Landau criterion looks for instability making nucleation of vortices possible. But the peak of the barrier, which suppresses phase slips, usually occurs at the later stage of the phase slip:  vortex expansion {\em after} its nucleation.\footnote{A unique example, when  the barrier peak for phase slips appears not at vortex expansion, but at the earlier stage of nucleation of a vortex core, is the spin-precession vortex in the $B$ phase of the superfluid  $^3$He\cite{Adv,Son08}}  At this stage the vortex is a macroscopic defect described by macroscopic hydrodynamics, while  the Landau criterion deals with elementary excitations. Nevertheless,  barriers for phase slips disappear at phase gradients of the order of values determined from the Landau criterion.\cite{EBS} Thus  the Landau criterion yields a qualitatively accurate upper bound on velocities at which superfluidity is still possible. 

An interesting feature of the spin-1 BEC is that  like superfluid $^3$He, it combines properties of a common  superfluid and of a magnetically ordered system. \cite{UedaR}. Correspondingly, one may expect coexistence and interplay of spin superfluidity and more common mass superfluidity. Investigation of this coexistence  is the main goal of the present work. Similarly to common ferromagnets, the spin-1 BEC is   described at macroscopical scales by the Landau--Lifshitz--Gilbert (LLG) theory\cite{LLstPh2} but extended by inclusion of an additional degree of freedom of fluid motion as a whole.\cite{Lam08}

The coexistence of two types of superfluidity has a profound effect on superfluid properties. The presence of spin degree of freedom (the multicomponentness boson wave function) essentially decreases critical velocity for mass superfluidity and even totally suppresses mass superfluidity if the spin space has full spherical symmetry. Uniaxial anisotropy in spin space is necessary not only for spin superfluidity, but for mass superfluidity as well. While the Landau critical velocity for mass superfluidity in single-component superfluids  is scaled by the sound-wave velocity and for spin superfluidity of localized spins is scaled by spin-wave velocity, in the case when  mass and spin superfluidity can coexist  the Landau critical velocities for both of them are determined by the lesser of two wave velocities. Usually this is the spin-wave velocity.  The Landau critical velocity for mass superfluidity is determined  by the spin-wave velocity even in the regime of the easy-axis anisotropy when spin currents vanish. This effect in multicomponent superfluids  was already known for the $A$ phase of $^3$He\cite{Bhat} and for  spinor  BEC of cold atoms,\cite{spin1/2,Duine} where it was also observed.\cite{ExpSp} This invalidates the assumption of some publications\cite{Flay}  that sound and spin wave velocities are  the Landau critical velocities for  mass and spin superfluidity  respectively ignoring the mutual effect of one on another at their coexistence. 

The present paper goes through  three essential steps (topology, Landau criterion, phase slips by vortices) in the investigation of spin and mass superfluidity in the ferromagnetic spin-1 BEC.  Our analysis shows that the most effective vortices for phase slips in the process of supercurrent decay, are low-energy nonsingular vortices with circulation of the spin phase, which at the same time  possess   circulation of the particle phase. Thus suppressing a pure spin supercurrent these vortices can produce  mass supercurrents, and vice versa. This makes complete relaxation to the ground state impossible. Complete relaxation of current states  is possible only with the participation of vortices of higher energy  with the same circulation of the spin phase but opposite circulation of the particle phase. The Landau criterion points out instability for nucleation of low-energy vortices. This leads to  an unexpected conclusion challenging previous common wisdom:  Sometimes the Landau criterion does not yield an upper bound for metastability of current states. At the phase transition ``easy plane--easy axis'', critical velocities at which metastable mass supercurrents  becomes impossible remain finite (we shall call them upper critical velocities),  while  the Landau critical velocities vanish. 
This restriction on using the Landau criterion of superfluidity is important for numerous examples of multicomponent superfluids and superconductors. 

\section{Multicomponent wave function in ferromagnetic spin-1 BEC}

The wave function  of bosons with spin 1 in the ferromagnetic state can be described by a complex vector  with three components $\psi_x$, $\psi_y$, and $\psi_z$,
\be
\bm \psi(\psi_x, \psi_y,\psi_z) ={\psi_0 \over \sqrt{2}} (\bm m+i \bm n),
        \ee{PW}  
where the scalar  $\psi_0$ and two mutually orthogonal unit  vectors $\bm m$ and $\bm n$ are real.  This vector wave function is an eigenfunction of a spin  along the axis defined by the unit vector  (see Ref.~\onlinecite{Ohmi1} and  problem 2 in Sec.~57 of \citet{LLqu})
\be
\bm s=-{i[\bm \psi^* \times \bm \psi]\over \psi_0^2}=\bm m \times \bm n.
        \ee{}
The two unit vectors $\bm m$ and $\bm n$ together with the third vector $\bm s$ form a triad.   Neutral and charged superfluids with such order parameters are called chiral superfluids. A similar order parameter wave function was used for description of the $A$-phase of $^3$He in the dipole-locked regime, but it corresponded to the Cooper pair with orbital moment 1 in the orbital space (with the unit  orbital vector $\bm l$ replacing the unit spin vector $\bm s$).\cite{VW} This presentation of the spin-1 wave function was called a Cartesian basis.\cite{UedaR}

One also can use another basis in the spin space called irreducible,\cite{UedaR} in which  three components $\psi_\pm,\psi_0$ of the vector  
\be
\bm \psi   =\left(\begin{array}{c}\psi_+ \\ \psi_0\\ \psi_-\end{array} \right)
     \ee{}
      are coefficients of the expansion of the wave function in eigenfunctions of the spin operator $\hat s_z$ with eigenvalues $s_z=\pm1,0$.\cite{,Ho98,UedaR}  The relations connecting $\psi_\pm,\psi_0$ with the components $\psi_x$, $\psi_y$, and $\psi_z$ of $ \bm \psi $ are  (apart from an arbitrary phase factor of unit modulus)
\be
\psi_x={\psi_+ -\psi_-\over \sqrt{2}},~~\psi_y={i(\psi_+ +\psi_-)\over \sqrt{2}},~~\psi_z=-\psi_0, 
    \ee{}
In the irreducible basis the components of the spin vector $\bm s$ are averaged values of the operators 
\bem
\hat s_x={1\over  \sqrt{2}}\left(\begin{array}{ccc} 0 & 1 & 0\\ 1 & 0& 1 \\ 0 & 1 & 0 \end{array} \right),~~\hat s_y={1\over  \sqrt{2}}\left(\begin{array}{ccc} 0 & -i & 0\\ i & 0& -i \\ 0 & i & 0 \end{array} \right),
\nonumber \\
\hat s_z=\left(\begin{array}{ccc} 1 & 0 & 0\\ 0 & 0& 0 \\ 0 & 0 & -1
\end{array} \right).
     \eem{}

It is important for hydrodynamics that the gauge transformation of the ferromagnetic spin-1 order parameter,
\bem \bm m+i \bm n ~\to~(\bm m+i \bm n)e^{i\theta}
\nonumber \\
= (\bm m \cos\theta- \bm n\sin\theta )+i(\bm m \sin\theta+ \bm n\cos\theta ),
       \eem{}
 is equivalent to rotation around the axis $\bm s$  by the angle $ \phi_s=-\theta$ and therefore is not an independent symmetry transformation. So the full point symmetry group of the order parameter in the ferromagnetic spin-1 BEC  is the group SO(3) of three-dimensional rotations.\cite{Ho98} The group is not Abelian, and the angle of rotation around any axis including the axis $\bm s$ depends on the path along which the transformation is performed. In particular, if we deal with the phase $\theta=-\phi_s$, a result of two small consecutive variations  $\delta_1$ and $\delta_2$ of $\theta$ depends on the order of their realizations:
 \be
 \delta_1 \delta_2 \theta-\delta_1 \delta_2 \theta=\bm s\cdot [\delta_1 \bm s \times \delta_2 \bm s].
              \ee{com12}
This means that phase $\theta$ is not well defined globally, although its infinitesimal variations still make sense and the quantum-mechanical definition of the superfluid velocity 
\be
\bm v_s={\hbar\over m}\bm \nabla \theta,
                                               \ee{9.1}
is valid. Here $m $ is the mass of a boson. Because of \eq{com12} variation of the superfluid velocity is determined not only by  variation of the phase $\theta$ itself but also by variation of the spin vector $\bm s$.  Namely, assuming $\delta_1 \to d$ and $\delta_2 \to \nabla_i $ in \eq{com12}, one obtains that
\be
dv_{si}={\hbar\over m} (\nabla_i d\theta+  [\nabla_i \bm s \times \bm s] \cdot d\bm s).
        \ee{noncom}
Moreover,  the superfluid velocity is not curl free. Relating $\delta_1$ and $\delta_2$ with two gradients $\nabla_1$ and $\nabla_2$ along two different directions ($x$ and $y$,  or $y$ and $z$, or $z$ and $x$),  \eq{com12} yields the Mermin--Ho relation \cite{Mer} between vorticity and spatial variation of $\bm s$: 
\be
\bm \nabla \times \bm v_s={\hbar\over 2m}\epsilon_{ikn} s_i \bm \nabla s_k \times \bm \nabla s_n.
                                              \ee{9.2}
This relation has a dramatic impact on hydrodynamics of chiral superfluids.

\begin{figure}[t]
\includegraphics[width=.5\textwidth]{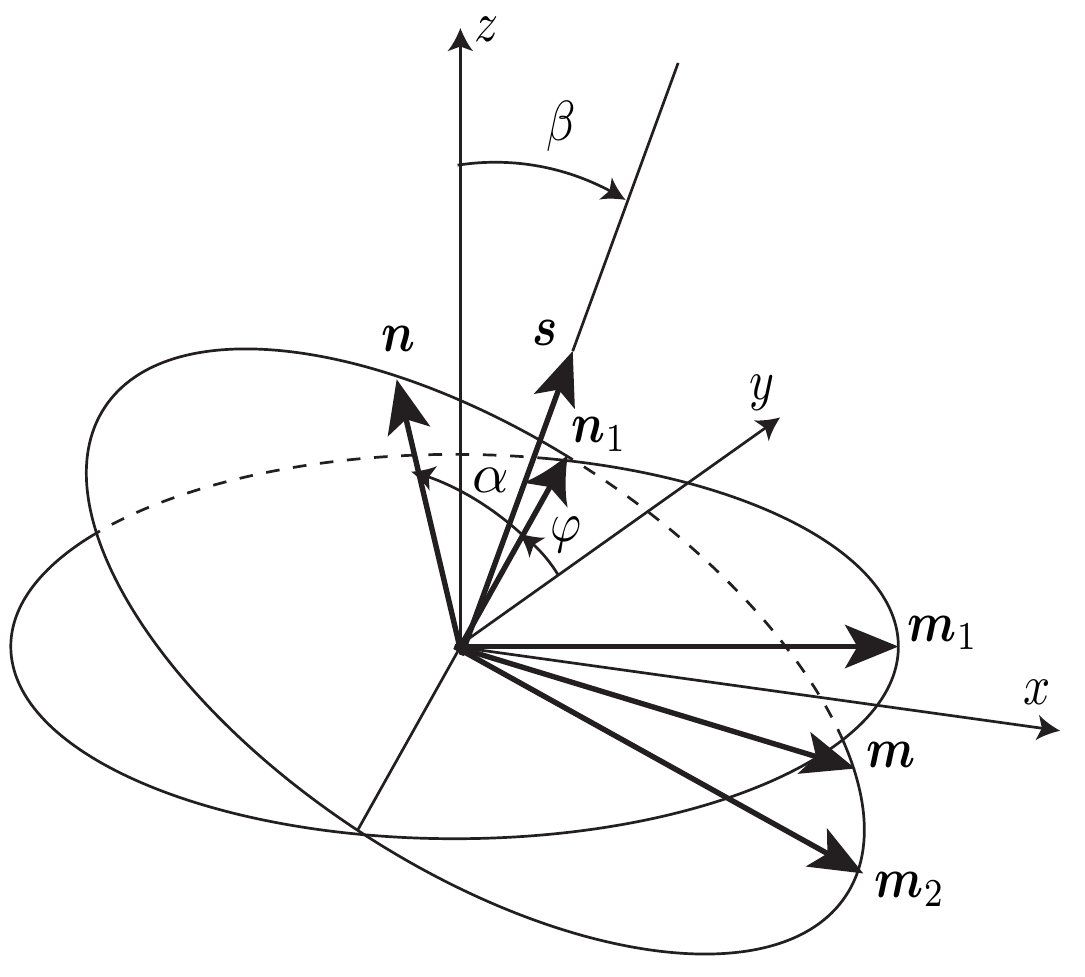}
\caption[]{Euler angles for the wave function triad. The original positions of $\bm m$, $\bm n$, and $\bm s$  are along the axes $x$, $y$, and $z$, respectively. The first rotation by the angle $\varphi$ is in the plane $xy$, which brings the first two vectors to the positions $\bm m_1$ and $\bm n_1$. The second rotation by the angle $\beta$ is in the plane confining the axis $z$ and the vector $\bm m_1$ (around the vector $\bm n_1$). This brings the vector $\bm s$ to its final position and rotates the vector $\bm m_1$ to $\bm m_2$. The last third rotation by the angle $\alpha$ is around the vector $\bm s$, which transforms the vectors $\bm m_2$ and $\bm n_1$ to the final vectors $\bm m$ and $\bm n$ determined by \eq{triaP}.}
\label{Fig1}
\end{figure}

It is possible to avoid  dealing with the globally undefined phase $\theta$  by introducing Euler angles as hydrodynamical variables. They determine  rotation of the triad  $\bm m,\bm n,\bm s$ with respect to the original triad $\hat x,\hat y, \hat z$ as shown in Fig.~\ref{Fig1}. The vector wave function in the Euler angles is
\be
\bm \psi =\left(\begin{array}{c}m_x+in_x \\ m_y+in_y\\ m_z+in_z\end{array} \right)={e^{-i \alpha}\over  \sqrt{2}}  \left(\begin{array}{c} \cos \beta  \cos \varphi -i\sin \varphi \\  \cos \beta \sin \varphi +i\cos\varphi \\ -\sin\beta
\end{array} \right),
   \ee{triaP}
in the Cartesian basis, and
\be
\bm\psi=e^{-i \alpha}\left(\begin{array}{c} e^{-i\varphi} \cos^2{\beta\over 2} \\  \sqrt{2} \sin {\beta\over 2} \cos {\beta\over 2} \\  e^{i\varphi} \sin^2{\beta\over 2}\end{array} \right)
      \ee{}
in the irreducible basis.\cite{Ho98} Independently from the basis, Cartesian components of the spin vector $\bm s$ are
\be
s_x=\sin\beta \cos \varphi, ~~ s_y =\sin\beta \sin \varphi, ~~ s_z =\cos \beta,
    \ee{tria}
and the phase ``gradient'' (it is not a true gradient!) $\bm \nabla \theta$, which determines the superfluid velocity in \eq{9.1}, is
\be
\bm \nabla \theta={m\bm v_s\over \hbar}=-\bm \nabla \alpha-\cos \beta \bm \nabla \varphi.
   \ee{vEu}
In contrast to the badly defined phase $\theta$, the Euler angles are well defined, and their gradients are curl free.
The Mermin--Ho relation in Euler angles becomes
\be
\left[\bm \nabla \times \bm v_s\right]={\hbar \over m}\sin\beta[ \bm \nabla \beta  \times \bm \nabla \varphi  ].  
  \ee{}

\section{Gross--Pitaevskii theory for chiral superfluids} 

Let us consider the extension of the Gross--Pitaevskii theory on a superfluid described by a vector wave function.\cite{Ho98,Ohmi}
The Lagrangian of the theory is 
\be
{\cal L}={i\hbar \over 2}\left(\bm \psi^*\cdot {\partial \bm \psi \over \partial t}-\bm \psi \cdot {\partial \bm \psi^* \over \partial t} \right) - H(\bm \psi,\bm \psi^*).
    \ee{Lag3}

For a Galilean invariant superfluid\footnote{A more general Gross--Pitaevskii theory for a superfluid without Galilean invariance was used for discussion of the intrinsic angular momentum in the $A$-phase of superfluid $^3$He at zero temperature\cite{Son84,EBS}} the  Hamiltonian   is
\bem
H ={\hbar^2 \over 2m} \nabla_i\psi_j^* \nabla_i\psi_j+{V|\bm \psi|^4\over 2},
      \eem{Ham3}
and the nonlinear Schr\"odinger equation   is
\be
i\hbar {\partial \bm \psi \over \partial t}={\delta  H(\bm \psi,\bm \psi^*)\over \delta \bm \psi^*}=-{\hbar^2  \nabla_j^2\bm \psi \over 2m}+V|\bm \psi|^2\psi.
    \ee{EqHam} 
Here we assume that  interaction is invariant with respect to rotations in the spin space and therefore the interaction energy $V|\bm \psi|^4/2$ depends only on particle density $|\bm \psi|^2$.  The  mass density  
\be
\rho =m\bm \psi^*\cdot \bm \psi
     \ee{}
      and  the mass current 
\bem 
j_i =-{i\hbar \over 2}(\psi_j^* \nabla_i \psi_j-\psi_j \nabla_i \psi_j^*)
   \eem{}      
are connected by  the mass continuity equation:
 \be
 {\partial \rho\over \partial t}+\bm\nabla\cdot \bm j=0.
     \ee{mC}
 
One can perform the generalized Madelung transformation,  after which  the vector wave function  $\bm \psi$ is described by the mass density $\rho=m \psi_0^2$, the spin vector $\bm s$, and the quantum-mechanical phase $\theta$, which determines the superfluid velocity in \eq{9.1}.      In the hydrodynamical approach usually they neglect dependence of the energy on density gradients (gradients of $\psi_0$) responsible for quantum pressure.\cite{EBS}  The Hamiltonian (\ref{Ham3}) after the Madelung transformation  becomes 
\bem
H = {\rho \over 2} v_s ^2+   {\hbar^2 \rho \over 4 m^2 }\nabla_i \bm s\cdot \nabla_i \bm s  +{V\rho^2\over 2m^2}.
      \eem{hamIn}
In hydrodynamical variables the generalized Gross--Pitaevskii theory yields the canonical equations of motion, two of which,  \eq{mC} and the Josephson equation for the phase $\theta$,
\be 
{\hbar \over m}{\partial \theta \over \partial t}+ \mu_0+{v_s^2\over 2}=0,
     \ee{JE}
are the same as in a nonchiral superfluid. Here 
\be
\mu_0 = {\hbar^2  \over 4 m^2 }\nabla_i \bm s\cdot \nabla_i \bm s  +V\rho
      \ee{mu}
is the chemical potential of the superfluid at rest. 

The third equation after the Madelung transformation of the Schr\"odinger equation (\ref{EqHam} ) is the equation for the unit vector $\bm s$,
 \be    
         S{\partial \bm s  \over \partial t}+  {\hbar\over m}(\bm j\cdot   \bm \nabla) \bm s +\left[ \bm s \times      {\delta H\over \delta \bm s} \right]     =0,  
    \ee{orbGP}
or taking into account the expression for the Hamiltonian (\ref{hamIn}),
\be    
         {\partial \bm s  \over \partial t}+ (\bm v_s\cdot   \bm \nabla) \bm s -{\hbar  \over 2 m \rho }\left[ \bm s \times      \nabla _i(\rho \nabla_i \bm s)\right]     =0.  
    \ee{orbGPa}
Here $S= \hbar \rho /m $ is the absolute value of the spin density vector 
\be
\bm S=S\bm s =i\hbar [\bm \psi \times \bm \psi^*].
       \ee{}
 For a fluid at rest ($\bm v_s=0$) \eq{orbGP} is identical to the LLG  equation for magnetization in a  ferromagnetic insulator. 
 
\Eq{orbGP} together with the mass continuity equation (\ref{mC}) gives the conservation law for the total spin:
  \be    
         {\partial  S_i  \over \partial t}+\nabla_jJ_{ij}   =0,  
    \ee{SpCon}
 where 
 \bem
 J^i_j =  S_i   v_{sj }   -\epsilon_{ikl}s_k {\partial H\over \partial \nabla_j s_l} 
 \nonumber \\
 = S_i   v_{sj }   -{\hbar^2\rho  \over 2 m^2 }\epsilon_{ikl}s_k \nabla_j s_l
       \eem{spCur}     
is the spin current tensor in the laboratory coordinate frame.   The first term in the expression   for the spin current presents advection of spin by fluid motion as a whole. This effect is trivial and has nothing to do with special conditions required for the existence of spin superfluidity. Only the second term,
\be 
 j^i_j =     -{\hbar^2\rho  \over 2 m^2 }\epsilon_{ikl}s_k \nabla_j s_l,
       \ee{spCurS}     
connected with stiffness of the spin texture will be later called the spin supercurrent. This is a spin current   in the coordinate frame moving with the superfluid velocity $\bm v_s$.    Using \eq{tria} connecting $\bm s$ with the Euler angles, the current of the $z$ component of spin is
\be
\bm j^z =-{\hbar^2 \rho\over 2m^2}\sin^2\beta  \bm \nabla\varphi.
    \ee{ssc}

      The functional derivative in \eq{orbGP}, 
\be
{\delta H\over \delta\bm s}={\partial  H\over \partial  \bm s} -\nabla_i {\partial  H\over \partial  \nabla_i  \bm s},
    \ee{}
was determined at fixed superfluid velocity $\bm v_s$. 
Meanwhile, variation of $\bm s$  produces also variation of $\bm v_s$. Bearing in mind \eq{noncom} connecting two variations  one can redefine the functional derivative with respect to $\bm s$ as
\be
{\tilde \delta H\over \tilde\delta\bm s}={\delta  H\over \delta  \bm s} +{\partial H\over \partial  v_{si}}{\hbar\over 4m}  [\nabla_i \bm s \times \bm s]={\delta  H\over \delta  \bm s} +{\hbar\over 4m}  [(\bm j\cdot \bm \nabla) \bm s \times \bm s].
        \ee{vsl}
Then \eq{orbGP} transforms to 
\be    
         S{\partial \bm s  \over \partial t}+\left[ \bm s \times      {\tilde\delta H\over \tilde \delta \bm s} \right]     =0. 
    \ee{orbGPt}

The Euler equation for the velocity $\bm v_s$ must follow from the Josephson equation (\ref{JE})  by applying  the gradient operator.
But one should take into account  the noncommutativity of the operators $\partial /\partial t$ and $\bm \nabla$ at their actions on the phase $\theta$. Namely, according to \eq{com12} 
 \be
\nabla_i  {\partial \theta\over\partial t}- {\partial (\nabla_i\theta)\over\partial t}
=\nabla_i  {\partial \theta\over\partial t}-{m\over \hbar} {\partial v_{si}\over\partial t}=\bm s\cdot \left[\nabla_i  \bm s \times {\partial \bm s\over\partial t} \right].
              \ee{}
After some algebra using the Mermin--Ho relation (\ref{9.2}) and the  equation (\ref{orbGP}) of spin dynamics one obtains the Euler equation 
\be
\dot {\bm v}_s+( \bm v_s \cdot \bm  \nabla) \bm v_s
+  \bm \nabla\mu_0+  {\hbar ^2  \over 2 m^2}\bm \nabla s_i {\nabla_j (\rho \nabla_j s_i )\over \rho}=0.
  \ee{}

\section{Uniaxial anisotropy}

Our equations derived from the Gross--Pitaevskii theory are isotropic in the spin space of the vector $\bm s$. But in an isotropic ferromagnet neither mass nor spin superfluidity is possible (see below).   Thus we shall add to our Hamiltonian terms breaking spherical symmetry but still invariant with respect to rotations around the axis $z$ (uniaxial anisotropy):
\be
H_A = - \gamma H_{ef} S s_z + {\rho Gs_z^2\over 2}.
         \ee{H_A}
Here $\gamma$ is the gyromagnetic ratio. The first term linear in $s_z$ is the Zeeman energy. The field $H_{ef}$ can be an external magnetic field but not necessarily. Processes violating the conservation law of spin usually are weak in comparison with the exchange interaction. By pumping magnons one can create a nonequilibrium  $z$ component of spin, which relaxes quite slowly, and this relaxation can be compensated by continuing magnon  pumping. With good accuracy one may consider  this state as a quasi-equilibrium state with fixed $z$ component of spin. Such states under the name magnon BEC were realized both in solids\cite{Dem6} and in ferromagnetic spin-1 BEC.\cite{MagBEC} Then $H_{ef}$ is a Lagrange multiplier, which determines the value of fixed total spin.\cite{Mur} The second term in \eq{H_A} is called in magnetism  easy-axis ($G<0$) or easy-plane ($G>0$) anisotropy. In the theory of cold atoms they call it the quadratic Zeeman energy.\cite{UedaR} The anisotropy energy determines two possible phases with the orientational phase transition between them. At  
$\gamma SH_{ef} > \rho G$ the energy is minimal at $s_z=1$ (easy-axis phase), while at $\gamma SH_{ef} < \rho  G$ the spin is confined in the plane parallel to the $xy$ plane and corresponding to $s_z=\gamma SH_{ef} /\rho G$ (easy-plane phase). 
Because in the easy-plane phase invariance with respect to rotations around the axis $z$ is spontaneously broken, it
is also called the broken-axisymmetry phase.\cite{UedaR}

\begin{figure}[b]
\includegraphics[width=.5\textwidth]{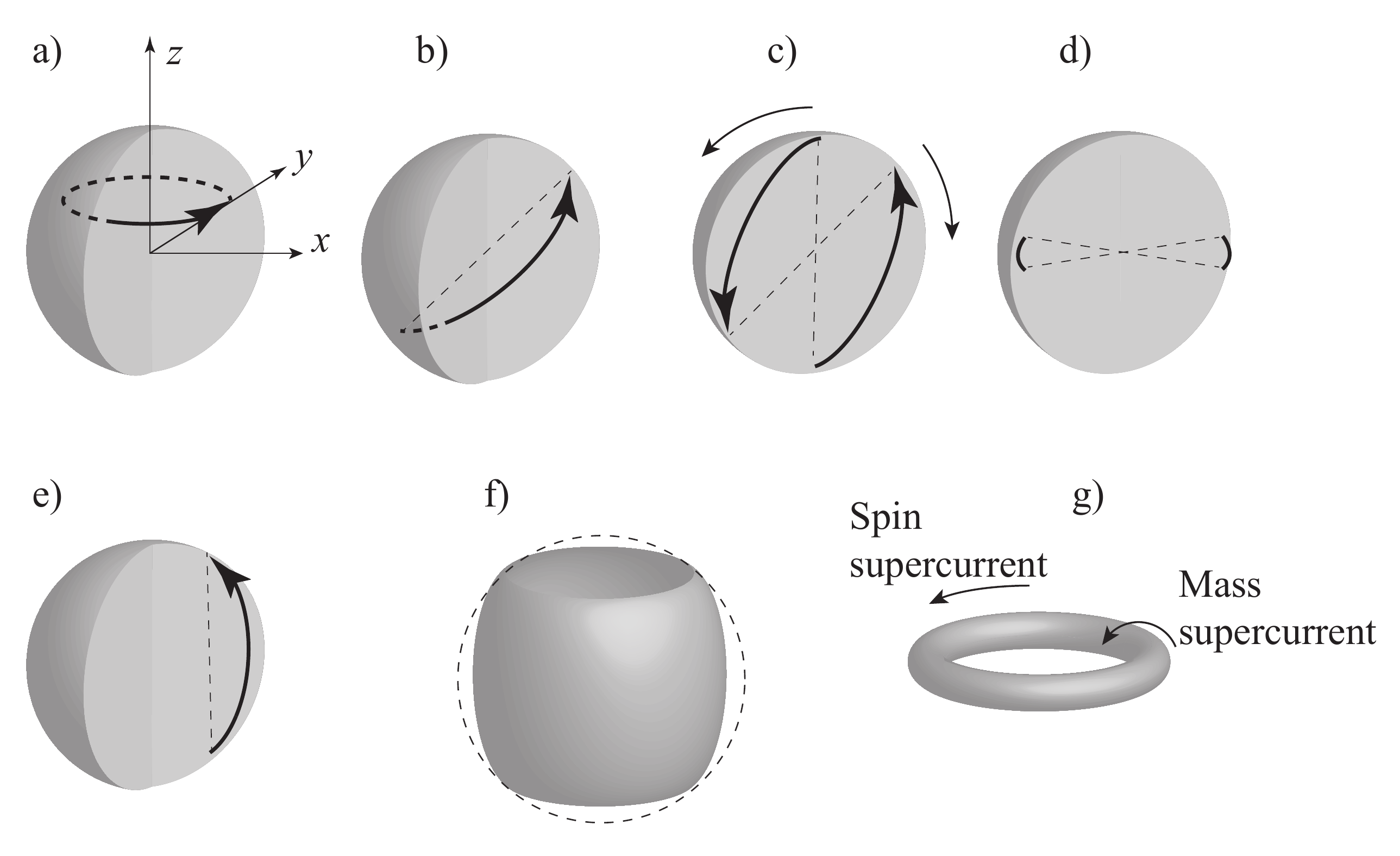}
\caption[]{Mapping of current states  on the order-parameter space SO(3). The pictures a)--e) refer to the case of spherical symmetry, while f) and g) take into account easy-plane anisotropy.
\newline
a) Spin current state maps on a close path in a plane parallel to the $xy$ plane. The path can be reduced by continuous deformation to the ground state [a point in the SO(3) sphere]. \newline
b) Mass current state maps on a path piercing the sphere once and connecting two equivalent antipodal points. This path cannot be deformed to a point and corresponds to a metastable single-quantum mass current. \newline
c) Mass current state maps on a path piercing the sphere twice. Thin arrowed lines show how the path must be deformed for reducing the path to a point [see d)]
\newline
d) The continuous deformation (homotopy) transforming the two-quanta path  to two antipodal points  equivalent one to another. The two-quanta  current (as any mass current with an even number of quanta) cannot be metastable. \newline
e) The  SO(3) sphere is twisted around the axis $z$ so that pairs of equivalent antipodal  points transform to two equivalent points symmetric one to another with respect to the $xy$ plane. \newline
f) The easy-plane anisotropy reduces the 3D filled SO(3) sphere to a 2D surface obtained by revolution of the path shown in e). Points at the upper edge of the surface are equivalent to points at the lower edge.\newline
g) The revolution surface is wrapped by connecting the upper and the lower edge. This transforms the order-parameter space to a torus. Spin and mass current states are expected to be metastable (persistent). }
\label{Fig2}
\end{figure}

Later in the paper we consider the case of incompressible liquid, when it is  enough to analyze only soft spin modes and to neglect density variation. 
Using the Euler angles  for the unit vector $\bm s$ as in \eq{tria},
the spin Hamiltonian including the anisotropy terms is 
\bem
H=  \rho  \left\{{v_s^2\over 2} + {\hbar ^2 \over  4m^2}\left[\sin^2\beta (\nabla \varphi)^2 +(\nabla \beta)^2\right]
\right. \nonumber \\ \left.
+{G(\cos\beta -s_0)^2\over 2}\right\},
   \eem{Hpol}
where 
\be
s_0={\gamma SH_{ef} \over \rho G}={\gamma \hbar H_{ef} \over m G},
    \ee{s0}
and the superfluid velocity $\bm v_s$ is given by \eq{vEu}.
The equations of spin dynamics in polar angles are
\be 
\dot \beta  +(\bm v_s\cdot \bm \nabla)  \beta=-{\hbar \sin \beta \over  2m } \nabla ^2 \varphi  -{\hbar  \cos \beta \over  m }\bm   \nabla  \varphi \cdot \bm \nabla \beta , 
     \ee{sde1}
\bem
\dot \varphi+(\bm v_s\cdot \bm \nabla)\varphi=  -{\hbar \over  2m}\left[ (\bm \nabla \varphi)^2\cos\beta -{\nabla ^2 \beta \over \sin\beta}
\right]
\nonumber \\
+{mG(\cos\beta -s_0) \over \hbar}.
         \eem{sde2}

\section{Topological analysis}

A qualitative prediction of possible mass or spin persistent current can be obtained from the analysis  of topology of the order-parameter space (vacuum manifold\cite{Man}) of the superfluid. It helps to know beforehand whether a medium is likely to have persistent supercurrents.
We start from the case of spherical symmetry in the spin space, when the order-parameter space is the space of group SO(3) of three-dimensional rotations of the order-parameter triad, as already mentioned in Sec.~\ref{Intr}.
 Each rotation of the triad through some angle  about the axis specified by the unit vector $\hat n$  maps on a point at some radius of the sphere parallel to $\hat n$ with distance of this point from the center equal to the rotation angle varying from 0 to $\pi$. Diametrically opposite (antipodal) points on the surface of the sphere correspond to the same state. This order-parameter space was already used in the past for the analysis of spin superfluidity in $^3$He.\cite{Son88} 

Current states in a closed ring channel map on closed paths in the order-parameter space. One may expect metastable persistent currents if  paths in the order-parameter space belong to topological classes different from the ground state (any point of the sphere). In other words, there is no continuous transformation (homotopy) reducing the path to a point. Spin current states with the nonzero circulation of the spin phase $\varphi$ map on circumferences parallel to the $xy$ plane, which are easily contracted to a point [Fig.~\ref{Fig2}a], so spin persistent currents are impossible. Mass current states map on paths piercing the sphere interior and  connecting equivalent antipodal points. A single-quantum current path piercing the sphere once is shown in Fig.~\ref{Fig2}b. One cannot reduce it to a point (the ground state). However, there is homotopy reducing  a two-quanta current path to the ground state shown in Figs.~\ref{Fig2}c and \ref{Fig2}d. This is true also for any current with an even number of quanta. Any mass current with an odd number of quanta belongs to the topological class of  single-quantum  currents.  Thus the homotopy group $\pi_1$ for the order-parameter space in the isotropic ferromagnetic spin-1 BEC is $Z_2$, i.e., consists from  two classes, and a single-quantum persistent mass current is possible. However, this is not a macroscopical persistent current because in a channel of macroscopical length a single-quantum persistent current corresponds to the fluid velocity inversely proportional to the channel length, which is negligible in the thermodynamic  limit. Thus in a multicomponent superfluid like a spherically symmetric ferromagnetic spin-1 BEC, or $^3$He-$A$ persistent currents, either of mass or spin, are impossible. This was known for $^3$He-$A$ long ago. 

In the presence of easy-plane anisotropy confining the spin vector $\bm s$ in a plane with $|s_z|<1$ the order-parameter space contracts the filled three-dimensional (3D) sphere to a tw0-dimensional (2D) surface. It is easier to picture this surface deforming (twisting) first the original SO(3) sphere as shown in Fig.~\ref{Fig2}e. After this deformation the 2D  space for the easy-plane order parameter is 
obtained by revolution of the path in Fig.~\ref{Fig2}e. This yields the 2D surface shown in Fig.~\ref{Fig2}f.  Wrapping of this surface by connecting the equivalent upper and lower edges transforms the order-parameter space to a surface of a torus  (Fig.~\ref{Fig2}g). It is evident that this topology allows both spin and mass macroscopic persistent supercurrents mapping on circumferences in cross sections of the torus by a vertical plane (mass supercurrents) or on paths around the axis $z$. All these paths belong to topological classes different from any ground state. Thus topology predicts the  possibility of both mass and spin persistent currents for the easy-plane anisotropy.

In the case of easy-axis anisotropy the order-parameter space is a vertical line connecting the northern and the southern poles of the sphere. Spin  superfluidity is out of question simply because the in-plane spin component $s_\perp$ vanishes in this case. But mass supercurrent states map on the path going along the line connecting the poles and piercing the sphere  many times. Whatever number of times the path is piercing the sphere, odd or even, there is no homotopy reducing this path to a point. Thus easy-axis anisotropy allows mass supercurrents.

\section{Collective modes and Landau criterion }

In the topological analysis it was supposed that gradients of phases (velocities) were very small, while at growing gradients one reaches the critical phase gradient values when the gradient kinetic energy becomes equal to the energy, which makes current states stable. The first estimation of the critical gradient (critical velocity) was done by Landau.  According to the Landau criterion, a supercurrent with velocity $\bm v_s$  is stable as far as any quasiparticle has a positive energy in the laboratory frame:
\be
\varepsilon=\varepsilon_0(\bm p) -\bm p \cdot \bm v_s >0,
        \ee{LCR}
where $\varepsilon(\bm p)$ and $\varepsilon_0(\bm p)$ are quasiparticle energies with the momentum $\bm p$ in the laboratory frame and the frame moving together with the fluid. For an isotropic spectrum this criterion is violated at the Landau critical velocity
\be
v_L =\mbox{min}{\varepsilon_0( p)\over p}.
     \ee{}
One can reformulate this as a condition imposed on the spectrum of classical collective modes with  frequency $\omega =\varepsilon /\hbar$ and wave number $k=p/\hbar$. The criterion can  be formulated as the condition $\omega(\bm k)>0$ for all collective modes in the laboratory frame with any possible wave vector $\bm k$.

Let us consider a uniform state with $s_z =\cos\beta$, the mass current proportional to the transport velocity $\bar{\bm  v}_s=-{\hbar \over m}(\bm   K_\alpha+\cos \beta \bm   K_\varphi)$ and the spin current (\ref{ssc}) determined by the constant gradients $\bm \nabla\varphi=\bm K_\varphi$ and $\bm \nabla \alpha=\bm K_\alpha $.
We linearize Eqs.~ (\ref{sde1}) and (\ref{sde2}) with respect to small perturbations $\bm v'_s$, $\bm \nabla \varphi'$, and $\beta'$  from the uniform state:  
\be 
\dot \beta'  +(\bm w\cdot \bm \nabla)  \beta'=-{\hbar \sin \beta \over  2m } \nabla ^2 \varphi' , 
     \ee{sdl1}
\bem
\dot \varphi'+(\bm w\cdot \bm \nabla)\varphi'+(\bm K_\varphi\cdot \bm v_s')
\nonumber \\
=-\sin\ \beta\left({mG\over \hbar} -{\hbar K_\varphi^2 \over  2m}  \right) \beta'  +{\hbar \over  2m}{\nabla ^2 \beta' \over \sin\beta}.
         \eem{sdl2}
Here  the velocity 
\be
\bm w=\bar{\bm v}_s+{\hbar  \cos \beta \over  m }\bm   K_\varphi={\hbar   \over  m }\bm   K_\alpha
     \ee{}
is determined by the gradient $\bm K_\alpha =\bm \nabla \alpha$. According to \eq{vEu}, a small perturbation of the superfluid velocity is
\be
\bm v_s '=-{\hbar\over m}[\bm \nabla \alpha' +\cos\beta \bm \nabla\varphi' -\bm K_\varphi \sin \beta \beta'  ].
   \ee{}

Now let us consider a plane-wave excitation, in which all perturbations are proportional to $e^{i\bm k \cdot \bm r-i\omega t}$.  
After exclusion of $\bm \nabla \alpha'$ from the linear equations of motion with the help of the incompressibility condition $\bm \nabla \cdot \bm v_s '=0$ the dispersion relation is
\bem
\left(\omega -\bm w \cdot \bm k\right)^2
\nonumber \\
={\hbar^2k^2\over 2 m^2}  \left[s_\perp^2 \left({m^2G\over \hbar^2}    +{K_\varphi^2 \over 2 }-{(\bm K_\varphi\cdot \bm k)^2 \over  k^2}\right)+{k ^2\over 2}  \right],
   \eem{dis}
where $s_\perp =\sin\beta$ is the in-plane component of the spin vector $\bm s$. The lowest threshold for instability is at small $\bm k \parallel \bm K_\varphi$ and zero frequency. The current state is stable  (Landau criterion) if
\be
w^2={\hbar^2 K_\alpha^2 \over m^2}<{s_\perp^2 \over 2} \left(G    -{\hbar^2K_\varphi^2 \over 2 m^2}\right).  
   \ee{LC}
 Neither mass, nor spin superfluidity is possible without easy-plane anisotropy $G>0$. 

If spin currents are absent ($K_\varphi=0$) the critical value of the transport superfluid velocity $\bar{\bm v}_s$ coincides with the spin-wave velocity
\be
v_c=s_\perp \sqrt{G\over 2}.   
     \ee{LCs}
The assumption of fluid incompressibility is valid as long as the sound velocity $c_s$ is much larger than the spin-wave velocity $v_c$. This result confirms that in the presence of the spin degree of freedom the critical velocity for mass superfluid is not sound velocity, but much smaller spin-wave velocity (see Introduction, Sec.~\ref{Intr}). 

According to inequality (\ref{LC}), in a resting  fluid ($\bar {\bm v}_s=0$) the critical gradient, at which the spin current loses stability, is
\be
K_c={2m \over \hbar}{s_\perp \sqrt{G}  \over \sqrt{1+3s_z^2}}.
       \ee{}
This differs by a numerical factor from the critical gradient $m v_c/\hbar$ connected with the spin-wave velocity $v_c$ in the absence of spin currents. Previous calculations of critical spin phase gradients\cite{Mur,Duine} ignored this difference, which is connected with  the effect of the phase gradient $\bm K_\varphi$ on the dispersion relation [the term $\propto K_\varphi^2$ in the right-hand side of \eq{dis}]. 

The collective mode, which  was used for derivation of the Landau criterion, is a Goldstone mode connected with broken axial symmetry in the easy-plane phase. Therefore it disappears in the critical point of the transition to the more symmetric easy-axis phase and does not exist after the transition. But one can check stability with respect to static fluctuations. In principle, this procedure yields the same threshold for instability as the analysis of the dynamic collective mode, as far as the latter shows that stability is lost in the zero frequency limit.\cite{Adv,Son17} In the easy-axis phase with $\beta=0$ there are mass supercurrents, which   become unstable in the critical point of the transition to the easy-plane phase. The critical point depends on phase gradients and is determined from the condition $\partial^2  H/ \partial \beta^2 =0$ at $\beta=0$: 
\bem
{\partial^2  H \over \partial \beta^2}=  \rho \left[G (s_0-1 )- { \hbar ^2\over m^2} \left(\bm K_ \alpha\cdot \bm K_\varphi  + { K_\varphi^2\over 2})\right)\right]
\nonumber \\
=  \rho \left[G (s_0-1 )-{ v_s^2\over 2}+ { \hbar ^2 K_ \alpha^2\over 2m^2}\right]=0 .
   \eem{stEA}
At $\beta=0$ the energy  depends only on the sum of the phase gradients  $\bm K_\varphi+\bm K_ \alpha$, but not from the two phase gradients separately. On the other hand, we look for the lowest threshold for instability. If so, we ignore  the term $\propto K_ \alpha^2$ in \eq{stEA}, which increases stability, and obtain the stability condition (Landau criterion)  for the easy-axis phase ($s_0>1$)
\be
v_s < \sqrt{2G(s_0-1)}.
     \ee{CLAx}

Thus critical gradients (critical velocities) vanish  at approaching the phase transition from above and from below. This seems natural for spin superfluidity because spin supercurrents are proportional to $s_\perp^2$ and vanish at  the phase transition and above it, whatever the spin phase gradients are. But  the impossibility of mass superfluidity at the transition does not look so evident. Indeed, zero $\bm w$ in the Landau criterion (\ref{LC}) does not mean zero  transport velocity $\bar{\bm v}_s$. At $w=0$ the criterion (\ref{LC}) becomes 
\be
0<{s_\perp^2 \over 2} \left(G    -{\hbar^2K_\varphi^2 \over 2 m^2}\right).  
   \ee{LCc}
This yields the critical velocity $v_c =\sqrt{G/2}$ which does not vanish at the phase transition, in contrast to $v_c$ given by \eq{LCs}.  But the physical meaning of this estimation also is not very transparent. The criterion points out when barriers providing metastability vanish--but what barriers? The right-hand side of the inequality (\ref{LCc}) is positive but its magnitude vanishes at the phase transition. Maybe barrier heights are also negligible?  This brings us to the question of what does really happen at reaching Landau critical values. 

In the case of  single-component superfluids the answer is well known. Supercurrents can decay only via phase slips, when vortices cross streamlines of the supercurrent.\cite{And6}  We shall discuss phase slips in Sec.~\ref{AT-S} after the analysis of vortex structure in Secs.~\ref{LLGv} and \ref{LLG-1}. The next section, Sec.~\ref{LLGv}, addresses  a simpler case of the LLG theory for localized spins when motion of the fluid as a whole is not possible. This prepares us for further analysis of the case of the ferromagnetic spin-1 BEC.

\section{Vortices in the LLG theory for localized spins} \label{LLGv}

The original LLG theory refers to a medium with localized carriers of spin when the degree of freedom of motion of the medium  as a whole is absent. So  we should delete in our equations everything, which is connected with the velocity $\bm v_s$.  Without anisotropy the order-parameter space is now $S_2$, i.e.,  a 2D surface of a unit  sphere in the 3D space. Every point of the surface corresponds to some direction of the unit vector $\bm s$.

\begin{figure}[t]
\includegraphics[width=0.5\textwidth]{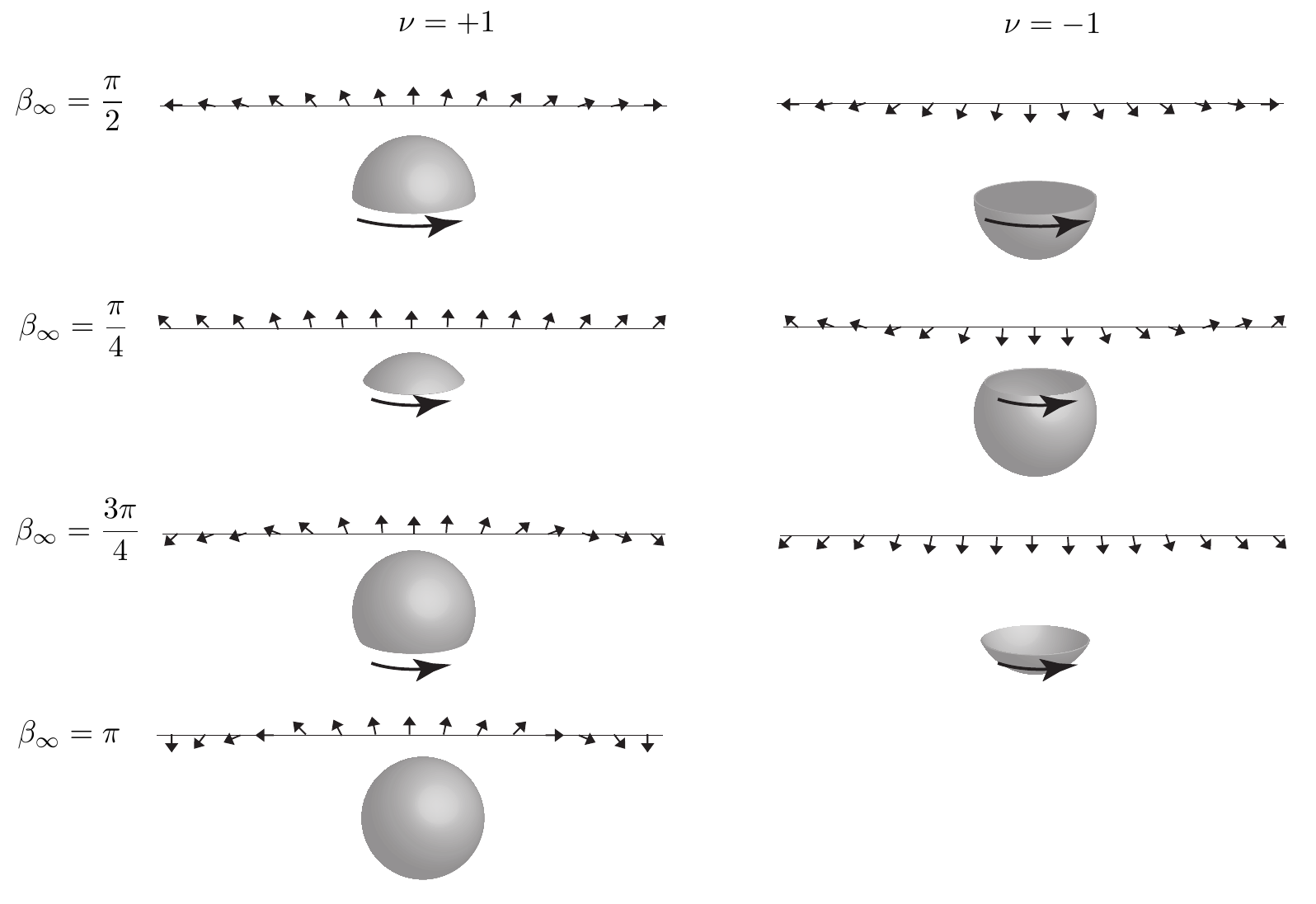}
\caption[]{ Spin vectors  $\bm s$ in axial cross-sections of skyrmion cores and mapping  on the space $S_2$ for vortex states with polarizations $\nu =\pm1$ and polar angles $\beta_\infty =\pi/4,~\pi/2,~3\pi/4$, and $\pi$. Larger arrows show direction of circular spin currents around the vortex (skyrmion) axis.}
\label{Fig3}
\end{figure}

Let us start from the easy-plane anisotropy case when the order-parameter space reduces to a circumference of the sphere $S_2$. The circumference corresponds to some fixed value of $s_z$ ($|s_z|<1$). However, only the periphery of the vortex very far from its axis can map on this circumference. The core of the vortex maps on an upper (northern) or lower (southern) part of the sphere and is characterized by two topological numbers.\cite{NikSon} The first one is the winding number, i.e., the number of rotations the spin makes on going around a vortex (the analog of the number of circulation quanta for a vortex in superfluid hydrodynamics). The second number, which can be called polarization, takes two values $\nu=\pm 1$. Two possible signs  correspond to a sign of the spin component $s_z$ at the vortex axis. We choose direction of the axis $z$ so that the in-plane spin component rotates counterclockwise around it. Mapping of vortex states with two polarizations and various values of the polar angle $\beta_\infty$ far from the vortex  are shown in Fig.~\ref{Fig3}.  The positive polarization corresponds to mapping on the northern part of the sphere, while the negative polarization points out mapping on the southern part. The vortex core  has a structure of a skyrmion. The skyrmion charge is a measure of wrapping of the spin vector around the sphere $S_2$ and equal to $Q=\sin^2 {\beta_\infty\over 2}$, so the vortex at $\beta_\infty=\pi$ has the unit charge $Q=1$. At $\beta_\infty=\pi/2$ when at periphery the spin is confined in the $xy$ plane, the core skyrmion  is a meron, or a half-skyrmion with  the skyrmion charge one-half.\cite{Man,braun} Other values of $\beta_\infty$ correspond to other fractional skyrmion charges. Thus in the  easy-plane anisotropy phase $\beta_\infty >0$ the skyrmion charge is not quantized and may vary continuously.

Skyrmions shown in Fig.~\ref{Fig3} are Neel skyrmions with nonzero magnetostatic charges proportional to
\be
\bm \nabla\cdot \bm s = {d\beta \over dr}+{\beta\over r}.
    \ee{} 
But rotation in the spin space around axis $z$ transforms skyrmions to Bloch skyrmions. Our model is invariant with respect to this rotation and ignores the magnetostatic  interaction.

Let us consider a straight axisymmetric vortex using two polar coordinates $r,\phi$ in a 2D configurational space. The spin angle $\varphi$ does not depend on the radial coordinate $r$ and for a single-quantum vortex is equal to the azimuthal angle $\phi$. The gradient of $\varphi$ has only the azimuthal component equal to $1/ r$.
The polar angle $\beta$ depends only on $r$. Then the energy density given by the Hamiltonian  (\ref{Hpol})  does not depend on the azimuthal angle $ \phi$:
\be
 H=  \rho  \left\{{\hbar ^2 \over  4m^2}\left[{\sin^2\beta \over r^2} +\left(d \beta\over dr\right)^2\right]
+{G(\cos\beta -s_0)^2\over 2}\right\}.
   \ee{HpolA}
The Euler--Lagrange equation for this Hamiltonian is
 \bem
{d ^2 \beta \over dr^2}+{1\over r}{d  \beta \over dr}-\sin\beta\left({ \cos\beta\over r^2} 
-{\cos\beta -s_0 \over \xi^2}\right)=0,
         \eem{beta}
 where
 \be
 \xi={\hbar \over m\sqrt{2G}},
      \ee{xi} 
and $s_0 = \cos \beta_\infty$ is the value of $s_z$ at large distances from the vortex axis. At small $r$  $\beta \propto r$, while at large $r$ $\beta$ approaches the equilibrium value $\beta_\infty$:
\be
\beta \approx \beta_\infty - {\xi^2\cos \beta_\infty \over r^2 \sin \beta_\infty}.
    \ee{betaAs} 
Note that  correction to the asymptotic  polar angle $\beta_\infty$ changes a sign at $\beta_\infty=\pi/2$.
One can define the core radius as a distance $r$ at which the correction to the asymptotic equilibrium value $\beta_\infty$ becomes comparable with $\beta_\infty$ itself. This yields the core radius of the order $r_c \sim \xi$  except for  very small 
$\beta_\infty $, when  \eq{beta} becomes
\bem
{d ^2 \beta \over dr^2}+{1\over r}{d  \beta \over dr}-{ \beta\over r^2} 
-{(\beta^2 -\beta_\infty^2)\beta \over 2\xi^2}.
         \eem{beta1}
This equation is identical to the Gross-Pitaevskii equation for radial distribution of the density of the vortex in a single-component superfluid. It shows that the core radius diverges at $\beta_\infty \to 0 $ as $r_c \sim \xi/\beta_\infty$.

\begin{figure}[t]
\includegraphics[width=0.45\textwidth]{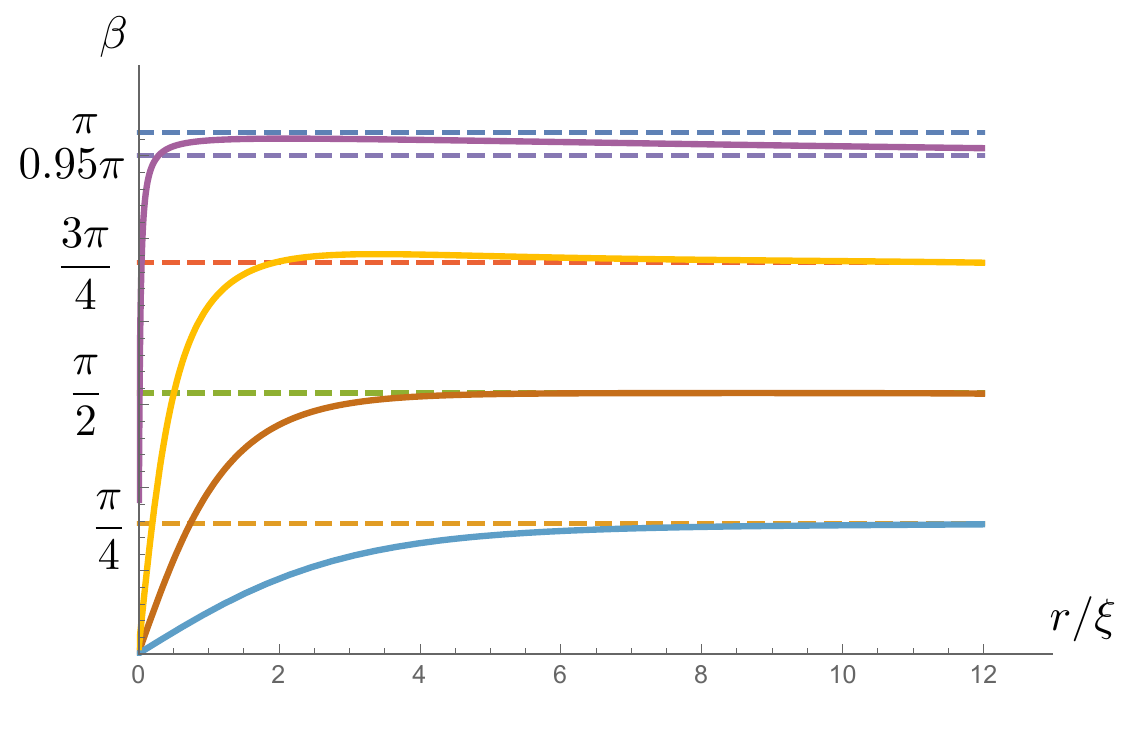}
\caption[]{ Plots  $\beta(r)$ for various $\beta_\infty$ shown by dashed lines and pointed out  along the ordinate axis.  The plots are solutions of \eq{beta} obtained in the LLG theory  for localized spins.}
\label{Fig4}
\end{figure}
 
In the easy-axis phase [$s_0>0$ in Eqs.~ (\ref{Hpol}) and (\ref{HpolA})] there is no spin vortices with circular spin currents at large distances. However, skyrmion as a topological defect is still possible as that with the charge $Q= 1$ shown in Fig.~\ref{Fig3} for $\beta_\infty =\pi$. But it cannot be stabilized at finite size. Without anisotropy ($\xi \to \infty $) spatial rescaling does not change the energy of the skyrmion and it can have any size. But the easy-axis anisotropy energy is smaller at smaller skyrmion size, and the skyrmion collapses to very small size. Its size can be stabilized by other interactions, e.g., by gradient terms of higher order,\cite{Ivan,Aban} or by the Dzyaloshinskii--Moriya interaction.\cite{Nagaos} 

Figure \ref{Fig4} shows plots of the polar angle $\beta$ as functions of the distance $r$ from the vortex axis for various values of the polar angle $\beta_\infty$.  These plots, as well as plots in Figs.~\ref{Fig5} and  \ref{Fig6}, were numerically calculated with  the standard program Mathematica.
At $\beta_\infty >\pi/2$ the curve $\beta(r) $ is not monotonic because of positive sign of correction to the asymptotic  polar angle $\beta_\infty$ in \eq{betaAs}. 
When $\beta_\infty$ approaches to the phase transition to the easy-axis phase at $\beta_\infty =\pi$ the core skyrmion size becomes very small as demonstrated by the curve for $\beta_\infty =0.95\pi$.  This is manifestation of instability of the $Q=1$	skyrmion with respect to the collapse mentioned above.

\section{Vortices in the ferromagnetic spin-1 BEC} \label{LLG-1}

In the ferromagnetic spin-1 BEC when the superfluid can move as a whole with the velocity $\bm v_s$, the spherical surface $S_2$ for the unit vector $\bm s$ is only a subspace in the larger space SO(3). But for our analysis of phase slips and vortices it is sufficient to consider only mapping on $S_2$.  
 
 Any vortex is characterized by two integers $N_\alpha$ and $N_\varphi$, which point out how many full $2\pi$ rotations the angles $\alpha $ and $\varphi$ perform along a path around the vortex. For an axisymmetric vortex
\be
  \bm \nabla \alpha = - N_\alpha {[\hat z \times \bm r]\over r^2} ,~~\bm \nabla \varphi =  N_\varphi  {[\hat z \times \bm r]\over r^2}, 
     \ee{gr}
and superfluid velocity $\bm v_s$ is equal to
\be
\bm v_v=  {\hbar[N_\alpha- N_\varphi\cos\beta(r)] [\hat z \times \bm r]\over mr^2}. 
    \ee{vgam} 
This velocity is not curl-free according to the Mermin--Ho theorem. Two singular contributions to the velocity at $r\to 0$  cancel one another if  $N_\alpha= N_\varphi$ at positive polarization [$\beta(0)=0$] or $N_\alpha= -N_\varphi$ at negative polarization [$\beta(0)=\pi$].  

Further we focus on a nonsingular vortex $N_\alpha= N_\varphi=1$ of positive polarization with the lowest energy.   
The velocity circulation $h(1-\cos\beta_\infty)/m$ at $r\to \infty$  for this vortex is equal to the double charge $2Q$ of the skyrmion in the vortex core.

\begin{figure}[t]
\includegraphics[width=0.45\textwidth]{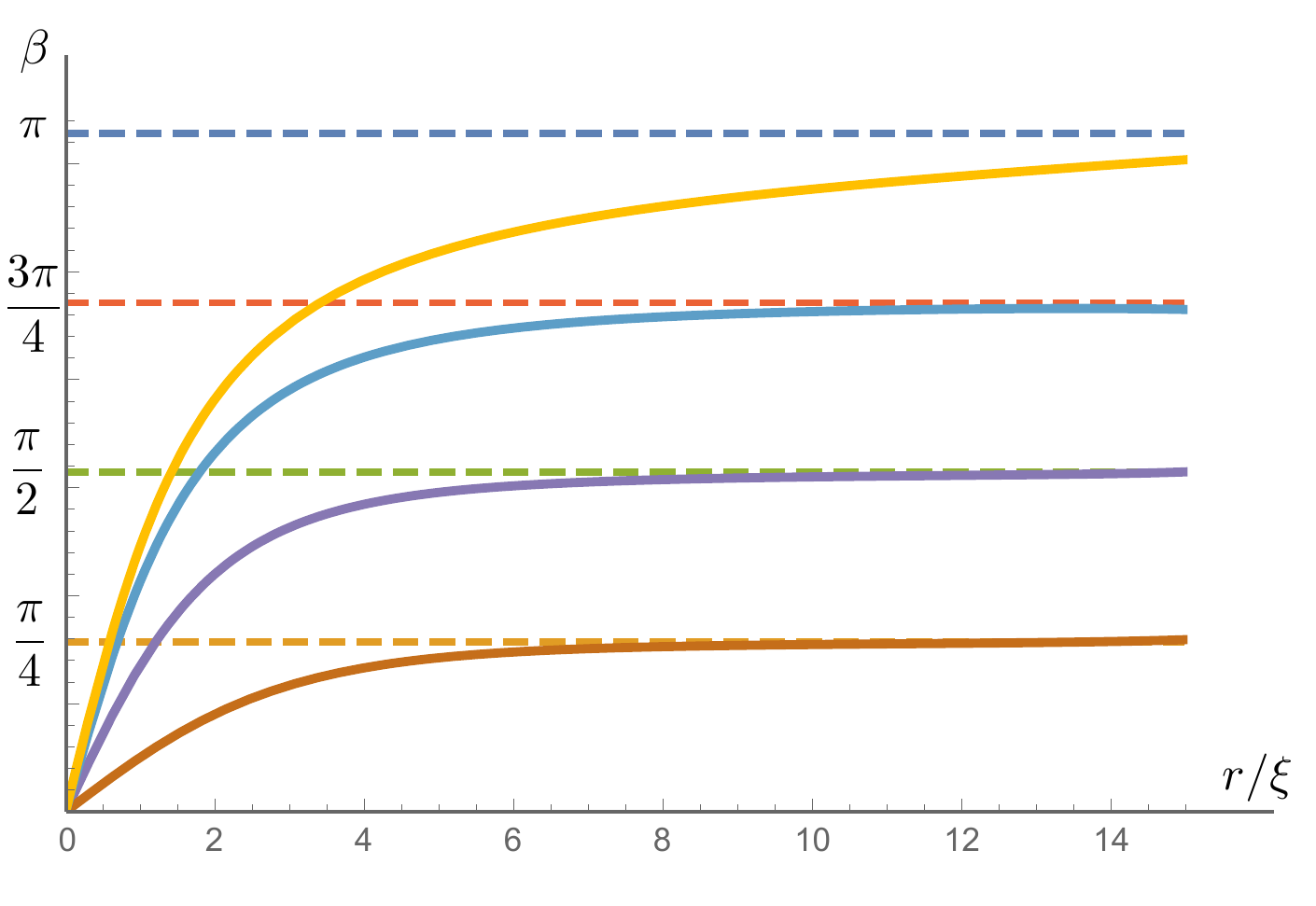}
\caption[]{ Plots  $\beta(r)$ for various $\beta_\infty$ shown by dashed lines and pointed out  along the ordinate axis. The plots are solutions of \eq{betaH} obtained in the theory  of the ferromagnetic spin-1 BEC.}
\label{Fig5}
\end{figure}

 Taking into account \eq{vgam}  the Hamiltonian (\ref{Hpol}) becomes
\bem
 H=  \rho  \left\{{\hbar ^2 \over  4m^2}\left[{(2-\cos\beta)^2-1 \over r^2} +\left(d \beta\over dr\right)^2\right]
\right. \nonumber \\ \left.
+{G(\cos\beta -s_0)^2\over 2}\right\}.
   \eem{HpolH}
The Euler--Lagrange equation for this Hamiltonian is
 \be
{d ^2 \beta \over dr^2}+{1\over r}{d  \beta \over dr}-\sin\beta\left({2- \cos\beta\over r^2} 
-{\cos\beta -s_0 \over \xi^2}\right)=0.
         \ee{betaH}
Figure \ref{Fig5} shows plots of the polar angle $\beta$ as functions of the distance $r$ from the vortex axis numerically calculated with \eq{betaH} for various values of the polar angle $\beta_\infty$.  In contrast to the LLG theory, the anisotropy is able to stabilize the skyrmion with charge 1 as the curve for $\beta_\infty =\pi$ shows: the size of the skyrmion core remains to be on the order of the length $\xi$. In the theory  of the A phase of superfluid $^3$He the vortex at $\beta_\infty =\pi/2$ (meron) was known as the Mermin--Ho vortex, while the vortex at $\beta_\infty =\pi$ was called the Anderson--Toulouse  vortex.\cite{Sal85} In the spin-1 BEC of cold atoms nonsingular vortices with skyrmion cores were observed experimentally.\cite{coreles}

\begin{figure}[b]
\includegraphics[width=0.45\textwidth]{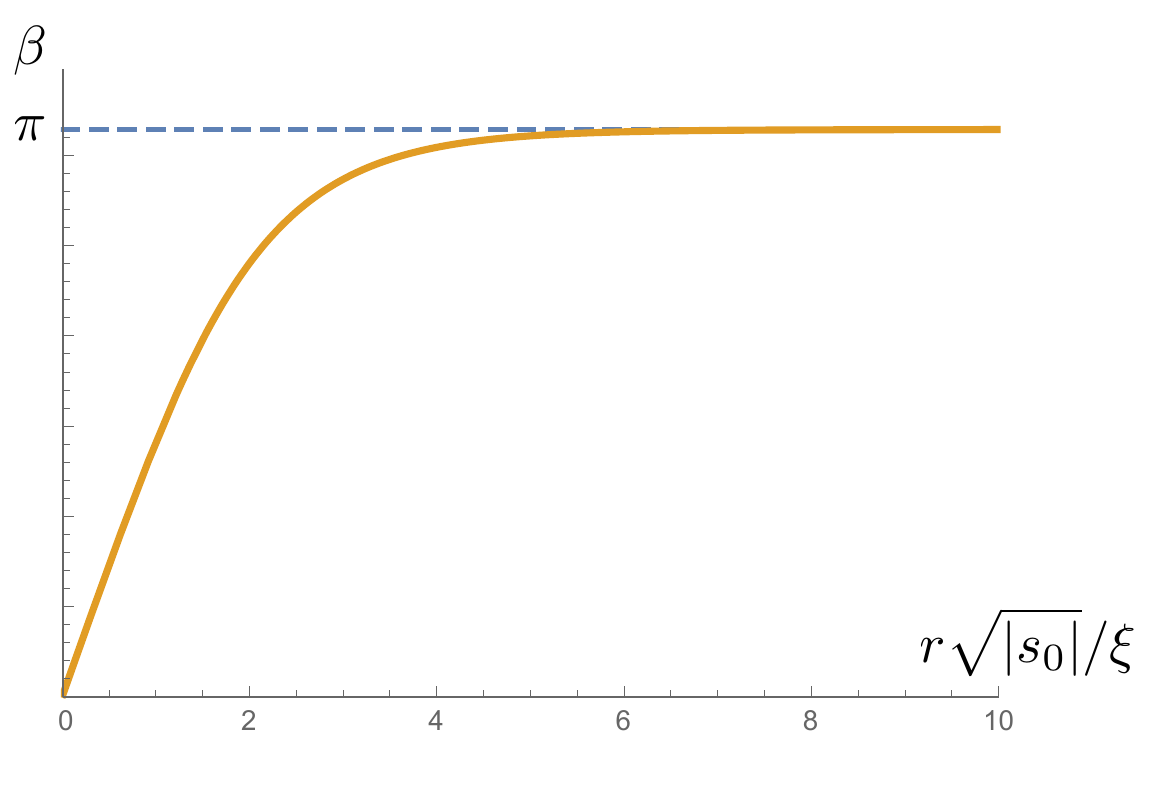}
\caption[]{ Plot  $\beta(r)$ for the Anderson--Toulouse  vortex (the core with the skyrmion of charge 1) at strong linear Zeeman effect when $|s_0| \gg 1$. The plot is  the solution of \eq{betaS}.}
\label{Fig6}
\end{figure}

In the easy-axis phase far from the transition when the linear Zeeman effect is strong ($|s_0| \gg 1$) the core radius drops. In this limit \eq{betaH}  becomes
\be
{d ^2 \beta \over dr^2}+{1\over r}{d  \beta \over dr}-\sin\beta\left({2- \cos\beta\over r^2} 
-{|s_0| \over \xi^2}\right)=0.
         \ee{betaS}
The numerical solution of this equation is shown in Fig.~\ref{Fig6}.
The spatial scale $\xi /\sqrt{|s_0| }$ of the equation  determines the core radius $r_c$. According  to Eqs.~ (\ref{s0}) and  (\ref{xi}), $r_c$ is about
\be
r_c \sim  \sqrt{\hbar \over m\gamma  H_{ef}}.
   \ee{}

\section{Phase slips and upper  critical velocity} \label{AT-S}

Let us find barriers suppressing phase slips. The first step of the phase slip is nucleation of a vortex ring in a 3D superfluid, or a vortex-antivortex pair in a 2D superfluid. The second stage is vortex expansion (growth of the vortex ring radius or of the distance between the vortex  and antivortex in a vortex-antivortex pair) in the direction normal to streamlines. The energetic barrier is the maximum energy at the second stage, i.e., at vortex expansion when the vortex can be described by hydrodynamics. In  single-component superfluids the vortex energy at the peak of the barrier is proportional to the logarithm $\ln(\hbar /m r_c \bar v_s )$.     Since the vortex energy was determined with logarithmic accuracy,  the core radius $r_c$ is known only by order of magnitude. In a single-component superfluid the core radius is of the order  $\hbar / mc_s$, where $c_s$ is the sound velocity. Thus the barrier  for vortex expansion disappears at the transport velocity   approximately equal to  the Landau critical velocity and one can rely on the Landau criterion at determination of the upper bound on the critical velocity. 

In  multicomponent superfluids there are various types of vortices. Let us start from the LLG theory for localized spins. It is evident that the energy of the skyrmion core is larger for a larger area of the mapping on the sphere $S_2$. Therefore, phase slips occur mostly via  vortices with the polarization $\nu=+1$ at $\beta_\infty <\pi/2$ and the polarization $\nu=-1$ at $\beta_\infty >\pi/2$. The phase gradient $\sim 1/r_c$ at which barriers for vortex expansion  vanish are of the same order as the critical gradient  obtained from the Landau criterion.\cite{Adv}

In the Galilean invariant ferromagnetic spin-1 BEC situation is more complicated. We focus on the vortex with single-quantum 
circulations of the spin phase $\varphi$ and of the angle $\alpha$, considered in Sec.~\ref{LLG-1}. There are mass and spin currents past the vortex determined by two gradients  $\bm\nabla\varphi=\bm K_\varphi $ and $\bm\nabla\alpha=\bm K_\alpha$. The total gradients and velocities are sums of fields induced by the vortex and currents:
\be
 \bm \nabla \alpha = - {[\hat z \times \bm r]\over r^2} +\bm K_\alpha,~~ \bm \nabla \varphi =  {[\hat z \times \bm r]\over r^2} +\bm K_\varphi.
       \ee{PSb}
Let us consider a straight vortex at distance $d$ from the wall. The gradients $\bm K_\alpha$ and $\bm K_\varphi $ must be parallel to the wall. By substituting \eq{PSb} into the Hamiltonian (\ref{Hpol})  and integrating over the 2D position vectors $\bm r$ one obtains for the vortex energy per unit length
\be
E_v=  {\pi \rho \hbar ^2(1 - \cos\beta_\infty )
 \over  m^2}\left[{3-\cos\beta_\infty \over 2}\ln {d\over r_c} 
- {2m \bar w d\over \hbar }\right],
   \ee{evor}
where the effective velocity
\be
\bar {\bm w}=-{\hbar\over m}\left( \bm K_ \alpha+ {1-\cos\beta_\infty  \over 2} \bm K_\varphi\right)
     \ee{}
was introduced. This velocity determines the  velocity of a vortex driven by mass and spin currents [see Eq.~(64) in Ref.~\onlinecite{Son18}]. Without spin currents ($K_\varphi=0$) the second term in \eq{evor} is $- \bar v_s P_v$ where $P_v$ is the vortex momentum.\cite{EBS} It appears at the Galilean transformation of the vortex energy to the moving coordinate frame.

This energy has a maximum at $d \sim \hbar/m \bar w$. The energy at the maximum is the height of the barrier, which with logarithmic accuracy is
\be
E_b=  {\pi \rho \hbar ^2[(2-\cos\beta)^2-1] \over  2m^2}
\ln {\hbar \over m \bar wr_c}. 
   \ee{barPS}
The barrier disappears at $\bar w_c \sim \hbar /m r_c$. If $\beta_\infty$ is not small the core radius $r_c$ is of the order of $\xi$, and the gradients (velocities) at which the phase slip barrier disappear are of the same order as predicted by the Landau criterion. The same is true for very small $\beta_\infty$, i.e., close to the phase transition, where $r_c \sim \xi/\beta_\infty$. This looks as if instability with respect to the phase slips starts at the same phase gradients as given by the Landau criterion. But this is not the end of the story.

As well as in the case of localized spins, phase slips are more probable with vortices with smaller areas of mapping on the spherical surface, which have smaller energies and larger $r_c$. But in  the spin-1 BEC there is a problem which is absent  in the case of localized spins.  The axisymmetric vortex with single quantum circulation of the spin in-plane angle $\varphi$ possesses also circulation of the mass velocity $\bm v_s$. Its sign coincides with $\varphi$ circulation for positive polarization, but opposite to it for negative polarization.
Thus, if we consider, e.g.,  the case of pure spin currents in  a resting superfluid, the phase slip with such a vortex decreases the spin current, but at the same time brings the superfluid into motion as a whole. Relaxation of the current state to the current-free ground  state is possible only if there are phase slips with vortices  having alternating signs of mass velocity circulation, i.e., alternating polarizations.  In other words, for complete current relaxation one needs phase slips  not only with vortices wrapping less than  a half of the surface $S_2$ with smaller energy, but also vortices wrapping more than  a half of $S_2$, which have larger energies. Apparently the Landau criterion  addresses instability only with respect to nucleation of vortices with less energy.

Let us find the state to which the uniform current state  can relax after phase slips via lower-energy (less than half of the skyrmion charge) vortices. If a phase slip with this vortex increases  the gradient $\bm \nabla \alpha$ it decreases the gradient $\bm \nabla \varphi$, or vice versa. Thus the gradient  energy in the process of relaxation is  (for the sake of simplicity we assume that the gradients $\bm \nabla \alpha$ and $\bm \nabla \varphi$ are parallel)
\bem
 H_\nabla=  {\rho\hbar ^2 \over  4m^2}  \{2[(\nabla \alpha_0-A)+\cos\beta (\nabla \varphi_0+A)]^2
\nonumber \\
+ \sin^2\beta (\nabla \varphi-A)^2 
\},
   \eem{HpolB}
where $\bm \nabla \alpha_0$ and $\bm \nabla \varphi_0$ are initial  gradients and $A$ is possible variation due to phase slips. 
Relaxation stops at the energy, which is minimal with respect to $A$. After this partial relaxation of the current state gradients satisfy the following relations: 
\bem
\nabla \alpha=\frac{1-\cos\beta }{2}\nabla\varphi,~~\bar v_s=-{\hbar\over m}\frac{1+\cos\beta }{2}\nabla\varphi.
   \eem{alpha}
At $\beta \to 0$   $\nabla \alpha=0$, and $\bar v_s=-(\hbar/ m)\nabla\varphi$. Thus in this  limit there are only mass supercurrents with the velocity fully determined by the angle $\varphi$ of rotation around the axis $z$. 

Final  relaxation of the current states to the ground state requires phase slips with vortices  wrapping more than a half of the surface $S_2$. These  vortices have negative polarization and induce the velocity field around them given by \eq{vgam} with  $N_\varphi =-N_\alpha=1$.  At $\beta_\infty \to 0$ these are  Anderson--Toulouse vortices with charge-1 skyrmion cores, and the height of the barrier is  
\be
E_{AT}={4\pi\rho \hbar^2 \over  m^2} \ln{\hbar \over m \bar wr_c}.
          \ee{AT}
Barriers  for expansion of these  vortices  also disappear  at gradients of the order of the inverse core radius $1/r_c$, but now $r_c$ is on the order of $\xi$, and the critical gradient   essentially exceeds the critical gradient $K_c \sim \beta_\infty/\xi$ following from the Landau criterion. So in the ferromagnetic spin-1 BEC near the phase transition the upper critical velocity for mass superfluidity is higher than the Landau critical velocity.

\section{Discussion and conclusions}

Let us summarize now conclusions on coexistence and interplay of mass and spin superfluidity in the ferromagnetic spin-1 BEC. Any superfluidity is possible only in the presence of uniaxial anisotropy  (the linear and quadratic Zeeman effect) in the spin space. In the condensate with  uniaxial anisotropy there are two phases with either easy-axis or easy-plane anisotropy. Spin supercurrents are possible only if there is an easy plane in which the spin is confined during its evolution  in space and time, but not possible at easy-axis anisotropy when spin is directed along the magnetic field  (the linear Zeeman effect is stronger than the quadratic Zeeman effect). On the other hand, mass superfluidity is possible in the both phases. 

We checked the Landau criterion for current states. The Landau criterion points out that the states with either mass or spin or both supercurrents become unstable at critical phase gradients, which are determined by anisotropy. Critical phase gradients vanish upon approaching the phase transition between the easy-plane and the easy-axis phases, as was shown previously.\cite{Duine} But  close to the transition instability of supercurrents overlaps with instability connected with the phase transition, and estimation of critical phase gradients (velocities) from the Landau criterion becomes unreliable. The instability threshold obtained from the Landau criterion triggers not a  complete decay of supercurrents but relaxation to another state with only  mass supercurrents without spin ones.  This conclusion was based on the analysis of the structure of vortices participating in phase slips in the ferromagnetic spin-1 BEC. 

These vortices have nonsingular cores with skyrmion structure. The  skyrmion charge is not quantized in the easy-plane phase and can have any fractal value. At any fixed value of the $z$  spin component $s_z$ there are 
 vortices with single quantum of spin phase of two types: with smaller skyrmion charge (less than 1/2)  and smaller energy and 
 with larger skyrmion charge (larger than 1/2)  and larger energy. The potential barrier for vortex expansion during phase slips is smaller for vortices with smaller energy. But they have circulations  of both  the particle and spin phase, and phase slips with these vortices are not sufficient for complete relaxation to the ground state. Complete relaxation to the ground state requires phase slips with vortices of larger skyrmion charge. This yields the upper critical velocity for mass superfluidity exceeding the Landau critical velocity.  It
  does not vanish and does not have any anomaly at the phase transition.
 
Our analysis was based on the assumption that nonsingular vortices participate in phase slips. Meanwhile, nonsingular vortices can be unstable with respect to disassociation to two singular vortices.\cite{MerminHoBEC,Loveg} This instability named dynamical instability was observed in spin-1 BEC experimentally.\cite{Shin} So our assumption requires justification.

For the sake of simplicity we consider the Anderson--Toulouse vortex in the easy-axis phase, where only mass superfluidity is possible. Generalization on other cases is straightforward. Let us compare the barrier (\ref{AT}) for the Anderson--Toulouse vortex and the barrier for a singular vortex. The velocity field around the singular vortex is given  by \eq{vgam} with $N_\alpha=1$ and $N_\varphi=0$. The energy of the vortex is\be
E_s={\rho h^2 \over 4\pi m^2} \ln{\hbar \over m \bar v _sr_c},
          \ee{}
where the core radius $r_c \sim \xi_0 =h /m c_s$ is determined by the sound velocity $c_s$, which is in our case much larger than the spin-wave velocity, and therefore $\xi \gg \xi_0$. The Anderson--Toulouse vortex is stable against disassociation  onto two single-quantum vortices  if $E_{AT} < 2E_s$.  But our problem is not disassociation of a pre-existing Anderson--Toulouse vortex. We look for an answer to the question of which vortex is easier to nucleate for realization of a phase slip. Our assumption  that nucleation of the   Anderson--Toulouse vortex is more probable is justified at a stricter condition $E_{AT} < E_s$, which yields an inequality
\be
\bar v_s > {\hbar \over m \xi}\left(\xi_0 \over \xi \right)^{1/3}.
     \ee{}
Thus close to  the upper critical velocity $\sim \hbar / m \xi$ mostly Anderson--Toulouse vortices participate in phase slips in agreement with our assumption. 

One may also expect that phase slips occur with nonsingular vortices, but with cores filled by the antiferromagnetic phase of the spin-1 BEC.\cite{Loveg} This option is ruled out if the energy difference between the ferromagnetic and the antiferromagnetic phase essentially exceeds the uniaxial anisotropy energy of the ferromagnetic phase.

In our analysis we assumed that supercurrents decay via phase slips  produced by vortex nucleation and subsequent vortex expansion. Meanwhile there were discussions in the literature about what happens at velocities exceeding  the Landau critical velocity $\bm v_L$ if vortex nucleation does not occur due to some reason. In Fermi superfluids occupation of single-particle levels with negative energies at supercritical velocities is restricted by the Pauli principle and therefore does not lead to complete decay of supercurrents. Instead, the superfluid density  starts to decrease and eventually vanishes at the velocity  exceeding  $v_L$ by a numerical factor of order unity.\cite{Bardin} In the Bose $^4$He  superfluid \citet{Pit84} showed that at the velocity exceeding $v_L$ a periodic structure appears and  a  metastable supercurrent is still possible. A similar scenario was considered  by \citet{Bay12} for a weakly interacting Bose gas with the excitation spectrum curving downward.  In Bose superfluids supercritical supercurrents were also revealed at velocities only slightly exceeding $v_L$. Thus the velocity $v_L$ remains to  be a reasonable estimation of the upper bound on possible nondissipative supercurrents in scalar superfluids. In contrast, in a multicomponent ferromagnetic spin-1 superfluid the Landau critical velocity vanishes at the phase transition between  the easy-plane and the easy-axis phases,  while the upper bound on possible velocities of metastable mass supercurrents remains finite and is on the order of spin-wave velocity far from the phase transition.

In  reality it is very difficult to reach the Landau critical velocity or the upper critical velocity  in experiments, moreover, in the supercritical regimes discussed above.\cite{Pit84} Phase slips start at subcritical velocities since barriers for them can be overcome by thermal activation or macroscopic quantum tunneling. So one can observe supercurrent relaxation at velocities less than  the Landau  critical velocity if the time of the experiment is long enough. This makes the very definition of the critical velocity rather ambiguous  and dependent on duration of observation of persistent currents.\cite{Var}   Calculation of real critical velocities requires a detailed dynamical analysis of  processes of thermal activation or macroscopic quantum tunneling through phase slip barriers,\cite{EBS} which is beyond the scope of the present investigation. But our calculation of the upper critical velocity is the first necessary step in this direction.

\end{document}